# Breakdown of Maxwell Garnett theory due to evanescent fields at deep-subwavelength scale

Ting Dong,[1] Jie Luo[2,4], Hongchen Chu[1], Xiang Xiong[1], and Yun Lai[1,3]

[1]*National Laboratory of Solid State Microstructures, School of Physics, and Collaborative Innovation Center of Advanced Microstructures, Nanjing University, Nanjing 210093, China*
[2]*School of Physical Science and Technology, Soochow University, Suzhou 215006, China*
[3]*e-mail: laiyun@nju.edu.cn*
[4]*e-mail: luojie@suda.edu.cn*

**Abstract:** Deep-subwavelength all-dielectric composite materials are believed to tightly obey the Maxwell Garnett effective medium theory. Here, we demonstrate that the Maxwell Garnett theory could break down due to evanescent fields in deep-subwavelength dielectric structures. By utilizing two- and three-dimensional dielectric composite materials with inhomogeneities at the scale of $\lambda/100$, we show that local evanescent fields generally occur nearby the dielectric inhomogeneities. When tiny absorptive constituents are placed there, the absorption and transmission of the whole composite will show strong dependence on the positions of the absorptive constituents. The Maxwell Garnett theory fails to predict such position-dependent characteristics, because it averages out the evanescent fields. By taking the distribution of the evanescent fields into consideration, we made a correction to the Maxwell Garnett theory, such that the position-dependent characteristics become predictable. We reveal not only the breakdown of the Maxwell Garnett theory, but also a unique phenomenon of "invisible" loss induced by the prohibition of electric fields at deep-subwavelength scales. Our work promises a route to control the macroscopic properties of composite materials without changing their composition, which is beyond the traditional Maxwell Garnett theory.

# 1. INTRODUCTION

Controlling the optical properties of photonic materials is a subject of great importance. One important approach to construct advanced optical materials is to build composite materials by intermixing two or more homogeneous constituents at deep-subwavelength scales. Generally, such a composite material can be homogenized as a continuous effective medium with uniform properties (Fig. 1(a)). In 1904, Maxwell Garnett developed a simple but immensely successful effective medium theory (EMT) [1]. Fields within the unit cell are considered local, and material properties are determined by the polarization and magnetization vectors. In this way, the Maxwell Garnett EMT gives the effective permittivity (or permeability) of the composite in terms of the permittivities (or permeabilities) and filling ratios of the individual constituents of the composite material. In this scenario, the macroscopic optical properties (e.g. reflection, transmission and absorption) of the composite material is insensitive to the details of the constituents (e.g. the position of each inclusion), as they are averaged out in the effective medium description [2-5].

Local homogenization is insufficient for the correct description of wave behaviors in special composite structures involving extremely large wave vectors or surface wave resonances even in the deep-subwavelength scale [6-13]. In such cases, the effective parameters become nonlocal or spatially dispersive, i.e. dependent on wave vectors [14, 15]. For example, strong nonlocality can be induced by surface plasmons in metal-dielectric structures even at the deep-subwavelength scale , leading to unusual effects including additional modes [8] and parabolic dispersions [7, 10].

On the other hand, deep-subwavelength all-dielectric composite materials, where surface wave resonances are not supported, are generally believed to tightly obey the Maxwell Garnett EMT. Interestingly, very recently, Sheinfux et al. [16] show the breakdown of EMT in deep-subwavelength all-dielectric multilayers, which has been numerically and experimentally demonstrated [17-27]. They found that the transmission through the multilayer structure depends strongly on nanoscale variations at the vicinity of the effective medium's critical angle for total internal reflection. Under this circumstance, the transmission spectra of the actual multilayer and its effective medium model become significantly different, because the effective medium approach cannot capture the microscopic evanescent and propagating waves in different dielectric layers and tunneling effects [16, 17]. These works are focused on the one-dimensional (1D) dielectric multilayers, in which evanescent waves occur only under large incident angles.

In two-dimensional (2D) and three-dimensional (3D) dielectric composite structures, the scenario is totally different. Inhomogeneity in 2D and 3D models always produces scattering fields, which contain both propagating waves (black arrows in Fig. 1(a)) and evanescent waves (red wavy arrows in Fig. 1(a)) [28]. Such evanescent waves produce rapidly varying evanescent fields nearby the interfaces of dielectric inhomogeneities even in deep-subwavelength structures. Since the evanescent fields occur in a very small area, and simultaneously increased and decreased at different locations, they are averaged out in the traditional Maxwell Garnett EMT.

In this work, we investigate 2D and 3D all-dielectric composite structures at the deep-subwavelength scale. We find that no matter how small the inhomogeneities are (e.g. even at a scale $<\lambda/100$, $\lambda$ is the wavelength in free space), evanescent fields will always occur nearby the interfaces of dielectric inhomogeneities. Because of these inevitable evanescent fields, Maxwell Garnett EMT breaks down when tiny absorptive constituents are placed into the system. We find that the absorption and transmission of the whole structure rely strongly on the positions of the tiny lossy inclusions as they could experience totally different local fields at different positions. Since the traditional Maxwell Garnett EMT averages out the evanescent fields, it fails to predict such position-dependent characteristics. By taking the evanescent waves into consideration, we have developed a correction to the Maxwell Garnett EMT, which can accurately predict the position-dependent transmission and absorption for both 2D and 3D models. Moreover, we predict an interesting phenomenon of "invisible" loss induced by the prohibition of electric fields that appears besides the epsilon-near-zero (ENZ) inclusions. Our work thus reveals a mechanism to control the bulk properties of photonic composite materials without changing the composition. This mechanism is beyond the description of the traditional EMT.

## 2. BREAKDOWN OF MAXWELL GARNETT THEORY AND THE CORRECTION

To begin with, we first consider a deep-subwavelength 2D model illustrated in Fig. 2(a) under the illumination of transverse-magnetic (TM, out-of-plane magnetic fields) polarized waves. Generally, incident waves will be scattered into various directions by different inclusions. Such scattered waves are usually considered to be propagating waves (black arrows). However, scattered evanescent waves (red wavy arrows) will also emerge if the sizes of inclusions are comparable or smaller than the wavelength, as sketched in Fig. 1(a). For visualization, we simulate the wave propagation in a dielectric composite consisting of a dielectric host (relative permittivity $\varepsilon_h = 2$, width $w$, height $h = 0.8w$) and three different dielectric inclusions by using the finite-element software COMSOL Multiphysics. The three inclusions are of elliptical, circular and square cross sections, and exhibit relative permittivities of 1, 5 and 3, respectively. Periodic boundary conditions are set on the upper and lower boundaries, and a TM-polarized plane wave with electric-field amplitude of 1V/m and wavelength of $\lambda = 125h$ is incident from the free space on the left side. Figures 1(b), 1(c) and 1(d), respectively, present the snapshots of $z$-component $E_z$, $x$-component $E_x$ and amplitude $|\mathbf{E}|$ of electric fields. There are clearly rapidly varying electric fields nearby the inclusions (dielectric interfaces). These evanescent waves are evanescent in the forward and backward directions (i.e. the $z$ direction), but capable of transferring energy flux from high-$\varepsilon$ regions to low-$\varepsilon$ regions along the perpendicular directions (i.e. the $x$ direction) [28]. They can be understood as a direct consequence of the electromagnetic field boundary conditions. Generally, at this deep-subwavelength scale, electric fields are enhanced in the low-$\varepsilon$ inclusions (the elliptical one), but weakened in the high-$\varepsilon$ inclusions (the circular and square ones), as seen in Fig. 1(d).

According to the Maxwell Garnett EMT, we can calculate the effective permittivity $\varepsilon_{eff}$ of a $d$-dimensional composite based on [3],

$$\frac{\varepsilon_{eff} - \varepsilon_h}{\varepsilon_{eff} + (d-1)\varepsilon_h} = \sum_i f_i \frac{\varepsilon_i - \varepsilon_h}{\varepsilon_i + (d-1)\varepsilon_h}, \qquad (1)$$

where $\varepsilon_i$ and $f_i$ are the relative permittivity and filling ratio of the $i$-th inclusion. In Fig. 1(e), we compare the transmittance through the actual composite (dots) and its effective medium model (lines) by assuming $N$ layers of unit cells along the propagation direction (i.e. the $z$ direction). The effective medium prediction matches very well with the simulation results. This indicates that the existence of evanescent waves in this configuration does not largely affect the validity of the EMT. This is understandable because the evanescent fields nearby the inclusions are simultaneously enhanced and weakened at different locations, which are usually averaged out in the EMT.

Nevertheless, the Maxwell Garnett EMT will become inaccurate and even break down completely when the composite contains tiny dissipative inclusions, whose sizes are comparable to or smaller than the decay length of the evanescent fields. When such tiny absorptive inclusions are placed close to the large inclusions, they will experience strongly enhanced or weakened fields instead of the averaged fields, and this will lead to position-dependent absorption and transmission characteristics. In terms of absorption, the deviation from the Maxwell Garnett EMT could be significantly large, as we shall demonstrate in the following.

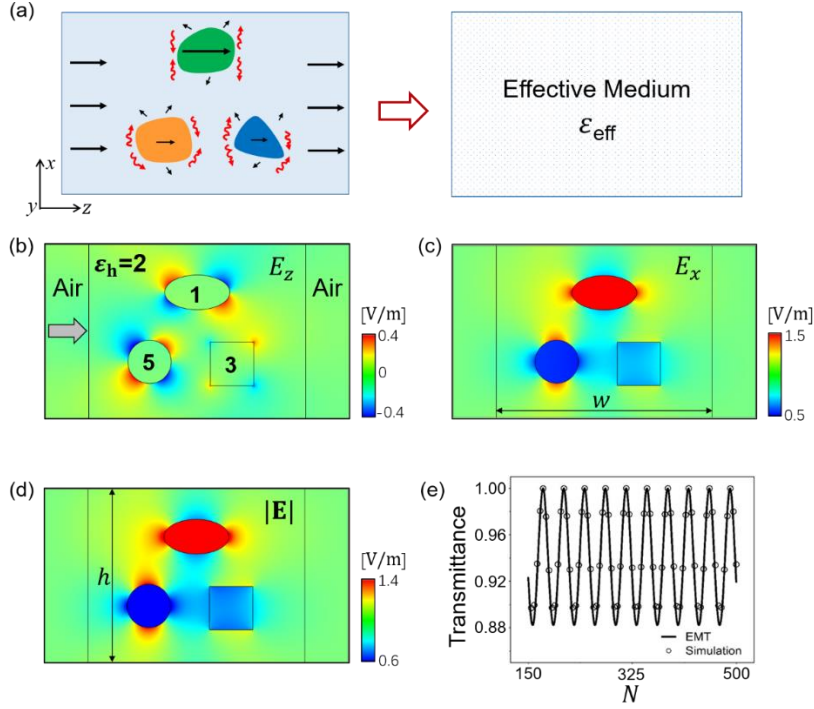

Fig. 1. (a) Schematic of a deep-subwavelength 2D all-dielectric composite structure (left), which generally can be treated as a continuous effective medium (right). [(b)-(d)] Snapshots of (b) $E_z$, (c) $E_x$ and (d) $|\mathbf{E}|$ when a TM-polarized wave is normally incident from the free space on the left side. The composite is composed of a host with $\varepsilon_h = 2$ and three inclusions with relative permittivities of 1, 5 and 3, respectively. The working wavelength is $\lambda = 125h$. (e) Transmittance through the actual composite (lines) and its effective medium model (dots) by assuming $N$ layers of unit cells along the propagation direction.

For simplicity, we consider an example of a 2D dielectric composite consisting of a dielectric host (relative permittivity $\varepsilon_h$, width $w$, height $h = 0.8w$) and an inclusion with circular cross section (relative permittivity $\varepsilon_1$, radius $r_1 = 0.15w$), as shown in Fig. 2. A TM-polarized wave with wavelength of $\lambda = 125h$ is incident from the free space on the left side. Figures 2(a) and 2(c) present the simulated $|\mathbf{E}|$-distributions for the case with $\varepsilon_h = 2$ and $\varepsilon_1 = 5$, and the case with $\varepsilon_h = 5$ and $\varepsilon_1 = 2$, respectively. In Fig. 2(a), it is seen that the electric fields on the upper/lower side of the inclusion (e.g. position 1) are enhanced, while those on the left/right side are weakened (e.g. positions 2 and 3). In Fig. 2(c), the situation is just the opposite. The existence of such evanescent fields is guaranteed by the electromagnetic field boundary conditions on the interface of the inclusions. In the following, we provide an understanding based on the quasi-static model. Since the inclusion is in the deep-subwavelength scale, we assume a background uniform electric field $\mathbf{E}_0$ (along the $x$ direction) in the host without inclusions, which is induced by the electric field of incidence in simulations. Then, the electric field inside the circular inclusion will be $\mathbf{E}_1 = \frac{2\varepsilon_h}{\varepsilon_h + \varepsilon_1}\mathbf{E}_0$ [29]. When $\varepsilon_1 > \varepsilon_h$ (or $\varepsilon_1 < \varepsilon_h$), we have $\mathbf{E}_1 < \mathbf{E}_0$ ($\mathbf{E}_1 > \mathbf{E}_0$), indicating weakened (or enhanced) electric fields inside the high-$\varepsilon$ (or low-$\varepsilon$) inclusions, as observed in Figs. 1(d), 2(a) and 2(c). Considering the continuity boundary condition at the host-inclusion interface, we find that the electric fields in the host nearby the interface are,

$$\mathbf{E}_h^{LR} = \mathbf{E}_1 = \frac{2\varepsilon_h}{\varepsilon_h + \varepsilon_1}\mathbf{E}_0 \text{ and } \mathbf{E}_h^{UL} = \frac{\varepsilon_1 \mathbf{E}_1}{\varepsilon_h} = \frac{2\varepsilon_1}{\varepsilon_h + \varepsilon_1}\mathbf{E}_0, \qquad (2)$$

on the left/right side (e.g. positions 2 and 3) and upper/lower side (e.g. position 1), respectively. Equation (2) implies $\mathbf{E}_h^{LR} < \mathbf{E}_0$ and $\mathbf{E}_h^{UL} > \mathbf{E}_0$ for the case of $\varepsilon_1 > \varepsilon_h$, while $\mathbf{E}_h^{LR} > \mathbf{E}_0$ and $\mathbf{E}_h^{UL} < \mathbf{E}_0$ for the case of $\varepsilon_1 < \varepsilon_h$, as observed in Figs. 2(a) and 2(c). In particular, if $\varepsilon_h \gg \varepsilon_1 \to 0$, we have $\mathbf{E}_h^{LR} = \mathbf{E}_1 \to 2\mathbf{E}_0$ and $\mathbf{E}_h^{UL} \to 0$, implying an interesting phenomena of the so-called "side scattering shadows" [30], as we will discuss in the following. If there exists an additional tiny absorptive inclusion close to the circular one, the absorptive constituent will experience dramatically different local fields at different positions. This would lead to position-dependent transmission characteristics as well as the breakdown of the traditional Maxwell Garnett EMT.

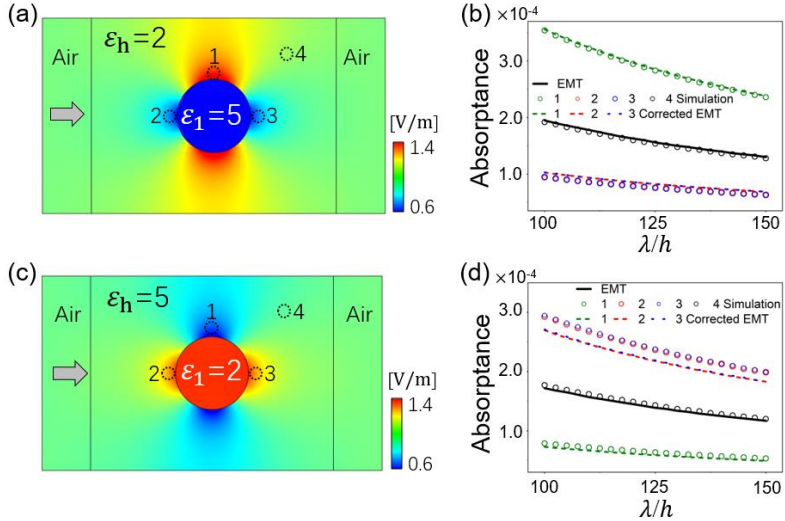

Fig. 2. [(a) and (c)] The $|\mathbf{E}|$-distributions for the composite structure with (a) $\varepsilon_h = 2$ and $\varepsilon_1 = 5$, (c) $\varepsilon_h = 5$ and $\varepsilon_1 = 2$ illuminated by a TM-polarized wave under normal incidence. The working wavelength is $\lambda = 125h$. The dashed circles denote the positions of additional tiny inclusions. [(b) and (d)] Absorptance by the composite with (b) $\varepsilon_h = 2$ and $\varepsilon_1 = 5$, (d) $\varepsilon_h = 5$ and $\varepsilon_1 = 2$ as functions of the working wavelength based on simulations of the actual composite (dots), traditional EMT (solid lines) and corrected EMT (dashed lines), when an additional tiny inclusion successively moves from position 1 to position 4. The radius of the tiny inclusion is $r_1/6$, and the relative permittivity is $2+i$ in (b) and $5+i$ in (d).

For demonstration, in Fig. 2(a), we add an additional dielectric inclusion with material loss (relative permittivity $\varepsilon_a = 2+i$, radius $r_a = r_1/6$), as illustrated by dashed circles at positions 1-4. The additional inclusion is much smaller than the original one, so that the original evanescent fields are not largely disturbed and the additional inclusion can experience enhanced or weakened local fields instead of the averaged fields. The absorptance by the composite is plotted in Fig. 2(b) as a function of the working wavelength when the additional inclusion successively moves from position 1 to position 4. We can see that the wave absorption for the additional inclusion at position 1 is much larger than that at other positions, because the electric field induced by evanescent scattering waves is largest at position 1. This clearly show the dependence of wave absorption on the positions of the tiny inclusion. However, the traditional EMT (i.e. Eq. (1)) ignores the evanescent fields and thus gives the same absorptance (black solid lines in Fig. 2(b)), which almost coincides with the result for the case of position 4 where evanescent waves almost disappear. The traditional EMT cannot capture the details of evanescent fields at the deep-subwavelength scale and thus fails in correctly describing such a position-dependent absorption.

Interestingly, by taking the evanescent fields into consideration, we can modify the formula of the traditional EMT, so as to give correct description of the position-dependent absorption and transmission characteristics. The effective permittivity of the $d$-dimensional composite containing $M$ large inclusions and $M'$ tiny inclusions can be calculated based on the corrected formula as,

$$\frac{\varepsilon_{eff} - \varepsilon_h}{\varepsilon_{eff} + (d-1)\varepsilon_h} = \sum_i^M f_i \frac{\varepsilon_i - \varepsilon_h}{\varepsilon_i + (d-1)\varepsilon_h} + \sum_j^{M'} f_{aj} \frac{\beta_j^2 (\varepsilon_{aj} - \varepsilon_h)}{d\varepsilon_h + \beta_j (\varepsilon_{aj} - \varepsilon_h)}, \qquad (3)$$

where $\beta_j$ denotes the correction factor for the $j$-th additional tiny inclusion (relative permittivity $\varepsilon_{aj}$, filling ratio $f_{aj}$). Here, $\beta_j$ can be evaluated by the ratio between the local field where the tiny inclusion is placed and the averaged field of the whole composite in the absence of additional inclusions through numerical simulations. The derivation of the corrected EMT is summarized in the Supplement 1. On the other hand, for simple cylindrical models, the value of $\beta_j$ can be roughly evaluated based on Eq. (2) as $\beta_j \approx |\mathbf{E}_h^j|/|\mathbf{E}_0|$ with $\mathbf{E}_h^j$ being the local electric field where the $j$-th tiny inclusion lies. In this way, we find that the $\beta_j$ varies in the range of $\frac{2\varepsilon_h}{\varepsilon_h + \varepsilon_1} \leq \beta_j \leq \frac{2\varepsilon_1}{\varepsilon_h + \varepsilon_1}$ when $\varepsilon_h < \varepsilon_1$ (or $\frac{2\varepsilon_1}{\varepsilon_h + \varepsilon_1} \leq \beta_j \leq \frac{2\varepsilon_h}{\varepsilon_h + \varepsilon_1}$ when $\varepsilon_1 < \varepsilon_h$). This indicates that the $\beta_j$ ranges from 0 to 2 in 2D models.

From the field-distribution in Fig. 2(a), we find that the correction factor $\beta$ at positions 1-4 is around 1.386, 0.7158, 0.7165 and 1.028, respectively. The $\beta$ at positions 2 and 3 is nearly the same, because the composite lies in the deep subwavelength scale. If the working wavelength tends to be infinitely long, we will get the exactly same $\beta$ under the electrostatic limit (see Eq. (2)). We also see that the $\beta$ at position 4 is near unity, indicating that the evanescent fields are mostly localized in very limited areas. Since Eq. (3) becomes Eq. (1) under $\beta = 1$, the absorption of the case at position 4 is close to the absorption predicted by the traditional EMT, as seen in Fig. 2(b). According to the corrected EMT (Eq. (3)), we plot the absorptance for the cases with the tiny inclusion at positions 1-3, as shown by the dashed lines in Fig. 2(b). We see that the results coincide with the simulation results quite well, demonstrating the validity of the proposed correction in the EMT.

We also re-analyze the composite with $\varepsilon_h = 5$ and $\varepsilon_1 = 2$ in Fig. 2(c) by adding an additional dielectric inclusion with material loss ($\varepsilon_a = 5+i$, $r_a = r_1/6$). The correction factor $\beta$ at positions 1-4 is found to be around 0.6486, 1.254, 1.257 and 0.9730, respectively. The absorptance of the composite as a function of the working wavelength is calculated based on the simulations of the actual composite, traditional EMT (Eq. (1)) and corrected EMT (Eq. (3)), as shown by the dots, solid lines and dashed lines in Fig. 2(d), respectively. We see that the traditional EMT fails to correctly describe the wave absorption when the additional tiny inclusion is close to the original large inclusion. Interestingly, the proposed corrected EMT can give an accurate description. We note that the deviation for the cases at positions 2 and 3 is mainly caused by the inhomogeneity of fields where the additional inclusion lies. If we further reduce the size of the additional inclusion and increase the wavelength, the deviation will become smaller.

The proposed correction of the EMT can be applied to dielectric composites containing arbitrarily-shaped inclusions. As an example, we consider a dielectric composite consisting of a dielectric host ($\varepsilon_h = 2$, width $w$, height $h = 0.8w$) embedded with two cloud-like inclusions ($\varepsilon_1 = 5$, $\varepsilon_2 = 1$) and three additional tiny inclusions (having the same relative $\varepsilon_a = 2+0.5i$ permittivity $\varepsilon_a = 2+0.5i$ and radius $r_a = 0.015w$) placed at positions 1-3, as illustrated in Fig. 3(a). Figure 3(b) presents the simulated $|\mathbf{E}|$-distribution in the absence of the additional inclusions under the illumination of a TM-polarized wave with $\lambda = 125h$, showing rapidly varying fields nearby the two cloud-like inclusions. Based on this field-distribution, the correction factors at positions 1-3 are evaluated as 1.246, 1.308 and 0.697, respectively. The absorptance of this composite is calculated as the function of wavelength based on the simulations of the actual composite, traditional EMT (Eq. (1)) and corrected EMT (Eq. (3)), as shown by the dots, solid lines and dashed lines in Fig. 3(c), respectively. The results clearly demonstrate the breakdown of the traditional EMT, and the validity of the corrected EMT in such a complex composite. In addition, in Fig. 3(d), we compare the transmittance through the effective media based on the traditional EMT (black lines) and the corrected EMT (red lines) by assuming $N$ layers of unit cells along the propagation direction. The results clearly show that the significant difference in absorption could lead to large deviation in the transmission for large samples. Here, we note that because the thickness of the composite sample ($\sim 100\lambda$) is much larger than the working wavelength, the interference of multiple reflections between interfaces is omitted in the calculation (see details in Supplement 1).

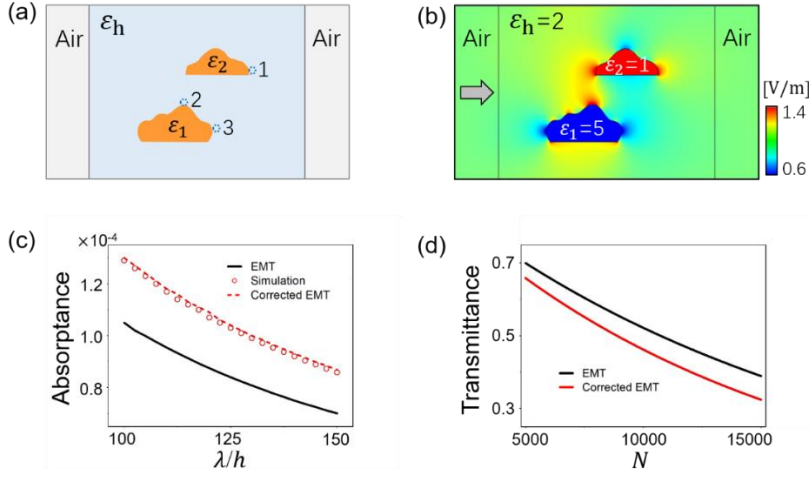

Fig. 3. (a) Illustration of a complex composite structure consisting of a host embedded with two cloud-like large inclusions and three tiny inclusions placed at positions 1-3. (b) The $|\mathbf{E}|$-distribution illuminated by a TM-polarized wave under normal incidence in the absence of the three tiny inclusions. (c) Absorptance by the composite as functions of the working wavelength based on simulations of the actual composite (dots), traditional EMT (solid lines) and corrected EMT (dashed lines). (d) Transmittance through the effective media based on the traditional EMT (black lines) and the corrected EMT (red lines) by assuming $N$ layers of unit cells along the propagation direction. The working wavelength is $\lambda = 125h$ in (b) and (d). The three tiny inclusions are the same with $\varepsilon_a = 2 + 0.5i$ and $r_a = 0.015w$. Other relevant parameters are $\varepsilon_h = 2$, $\varepsilon_1 = 5$ and $\varepsilon_2 = 1$.

## 3. "INVISIBLE" LOSS INDUCED BY EVANESCENT FIELDS

The position-dependent characteristic can lead to an interesting phenomenon, i.e. the disappearance of absorption in dielectric composites with absorptive constituents. This is impossible from the viewpoint of traditional EMT, because all inclusions contribute to the effective parameters. Interestingly, we find that when tiny absorptive inclusions are placed where $\beta \to 0$, the waves cannot "see" them. As a result, the effective permittivity of the whole composite will be totally independent of these tiny absorptive inclusions (see Eq. (3)), thus leading to the disappearance of total absorption. Equation (2) indicates that the area with $\beta \to 0$ can actually be obtained if the composite contains ENZ inclusions, which is also denoted as "side scattering shadows" [30].

As an example, we consider a 2D composite consisting of a dielectric host ($\varepsilon_h = 2$, width $w$, height $h = 0.8w$) embedded with a rectangular ENZ inclusion ($\varepsilon_1 = 0.001$, width $0.7w$, height $0.1w$) and two slabs of lossy dielectric ($\varepsilon_a = 2 + i$, width $0.35w$, height $0.01w$) coated on the upper and lower surfaces of the ENZ inclusion, as sketched in Fig. 4(a). Figure 4(b) displays the simulated the $|\mathbf{E}|$-distribution in the absence of the two lossy inclusions when a TM-polarized wave with $\lambda = 125h$ is normally incident from the free space on the left side. Clearly, due to the matching of the displacement fields inside and outside the ENZ medium, the electric fields becomes extremely small on the upper and lower sides of the ENZ inclusion. From the average field in the area where the lossy inclusions lie, we evaluate the correction factor $\beta$ as 0.151, significantly smaller than unity. Figure 4(c) presents the absorptance of this composite as a function of the working wavelength. As expected, the wave absorption in the actual composite (dots) is negligibly small, which is well predicted by the corrected EMT (dashed lines). The absorption predicted by the traditional EMT (solid lines) shows an obvious deviation from the actual absorption. Moreover, we plot the transmittance through the effective media based on the traditional EMT (black lines) and the corrected EMT (red lines) with $N$ layers of unit cells in Fig. 4(d). We can see that the transmission in both cases oscillates over $N$ as the result of Fabry-Pérot resonances. The transmission predicted by the traditional EMT (black lines) decreases quickly when $N$ increases. While the transmission oscillation predicted by the corrected EMT remains almost unchanged (red lines), indicating the disappearance of loss.

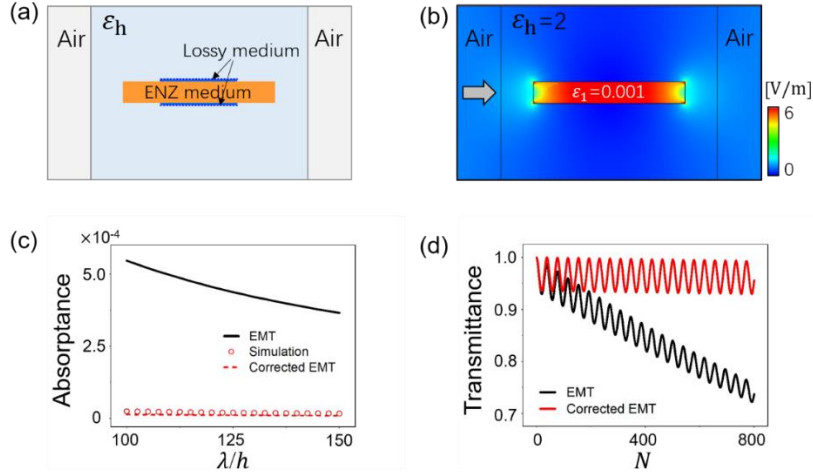

Fig. 4. (a) Illustration of a composite structure consisting of a host embedded with ENZ inclusion and two lossy inclusions. (b) The $|\mathbf{E}|$-distribution illuminated by a TM-polarized wave under normal incidence in the absence of the lossy inclusions. (c) Absorptance by the composite as functions of the working wavelength based on simulations of the actual composite (dots), traditional EMT (solid lines) and corrected EMT (dashed lines). (d) Transmittance through the effective media based on the traditional EMT (black lines) and the corrected EMT (red lines) by assuming $N$ layers of unit cells along the propagation direction. The working wavelength is $\lambda = 125h$ in (b) and (d). The two tiny inclusions are the same with $\varepsilon_a = 2 + i$. Other relevant parameters are $\varepsilon_h = 2$ and $\varepsilon_1 = 0.001$.

## 4. 3D MODELS

Besides 2D models, the breakdown of traditional Maxwell Garnett EMT also applies to 3D models. Figure 5(a) shows a practical 3D model consisting of a sphere of silicon (Si, $\varepsilon_1 = 12$, radius 15nm) and eight spheres of indium tin oxide (ITO, radius 3nm) in a host of silica (SiO$_2$, $\varepsilon_h = 2.1$, side length 50nm). The eight ITO spheres are evenly distributed along the equator line of the Si sphere. The relative permittivity of the ITO sphere is $\varepsilon(\omega) = \varepsilon_\infty - \dfrac{\omega_p^2}{\omega(\omega + i\Gamma)}$ [31], where $\varepsilon_\infty = 3.94$, $\omega_p = \sqrt{5.97} \times 10^{15}$ Hz, $\Gamma = 1.88 \times 10^{14}$ Hz, and $\omega$ is the angular frequency. A plane wave with electric fields polarized along the $x$ direction is normally incident from the free space on the left side. Figure 5(b) presents the simulated $|\mathbf{E}|$

-distribution in the absence of the ITO spheres when the working wavelength is 1400nm, showing enhanced fields nearby the poles and weakened fields nearby the equator line of the Si sphere. Similar to Eq. (2) for the 2D model, we can also evaluate the electric fields in this 3D model under the electrostatic limit as [29],

$$\mathbf{E}_h^{equator} = \frac{3\varepsilon_h}{2\varepsilon_h + \varepsilon_1}\mathbf{E}_0 \text{ and } \mathbf{E}_h^{pole} = \frac{3\varepsilon_1}{2\varepsilon_h + \varepsilon_1}\mathbf{E}_0, \qquad (4)$$

where $\mathbf{E}_h^{equator}$ and $\mathbf{E}_h^{pole}$ are, respectively, the electric fields in the host nearby the equator line and pole of the Si sphere. Equation (4) provides us a way to roughly evaluate the correction factor $\beta_j$ as $\beta_j \approx |\mathbf{E}_h^j|/|\mathbf{E}_0|$ in the simple spherical model. We find that the $\beta_j$ varies in the range of $\frac{3\varepsilon_h}{2\varepsilon_h + \varepsilon_1} \leq \beta_j \leq \frac{3\varepsilon_1}{2\varepsilon_h + \varepsilon_1}$ when $\varepsilon_h < \varepsilon_1$ (or $\frac{3\varepsilon_1}{2\varepsilon_h + \varepsilon_1} \leq \beta_j \leq \frac{3\varepsilon_h}{2\varepsilon_h + \varepsilon_1}$ when $\varepsilon_1 < \varepsilon_h$). This means that the $\beta_j$ ranges from 0 to 1.5 in 3D models.

Since $\varepsilon_1 > \varepsilon_h$ in the model in Fig. 5(a), we have $\mathbf{E}_h^{equator} < \mathbf{E}_0$ and $\mathbf{E}_h^{pole} > \mathbf{E}_0$, as observed in Fig. 5(b). This indicates the rapid variation of electric fields nearby the Si sphere, i.e. the existence of evanescent fields. Likewise, if there exist additional tiny inclusions experiencing the evanescent local fields, the traditional Maxwell Garnett EMT will fail to describe such location-dependent situations.

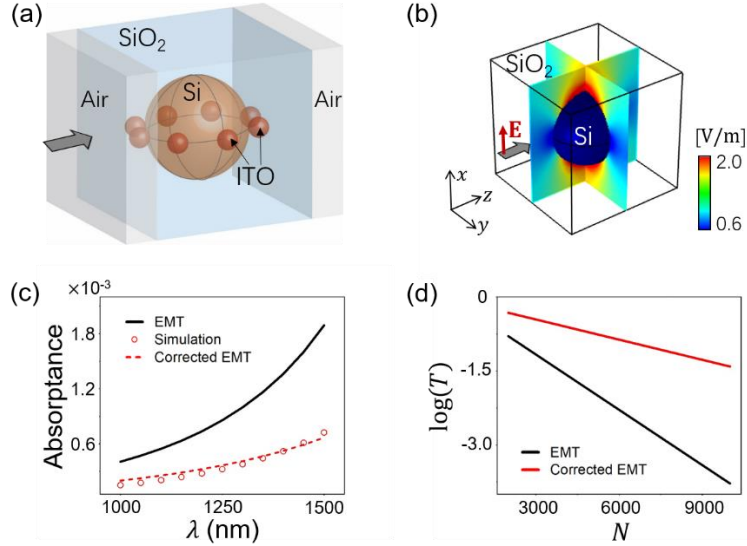

Fig. 5. (a) Illustration of a 3D composite structure consisting of a SiO2 host and a Si sphere (radius 15nm) surrounded by eight tiny ITO spheres (radius 3nm). (b) The $|\mathbf{E}|$-distribution illuminated by a plane wave under normal incidence in the absence of the ITO spheres. (c) Absorptance by the composite as a function of working wavelength based on simulations of the actual composite (dots), traditional EMT (solid lines) and corrected EMT (dashed lines). (d) Transmittance on a log scale, i.e. $\log(T)$, through the effective media based on the traditional EMT (black lines) and the corrected EMT (red lines) by assuming $N$ layers of unit cells along the propagation direction. The working wavelength is 1400nm in (b) and (d).

For verification, we have calculated the absorptance as a function of the working wavelength, as shown in Fig. 5(c). The dots and solid lines denote the results obtained through the simulation of the actual composite and the traditional EMT (Eq. (1)), respectively. Clearly, the effective medium model overestimates the absorption because the fields experienced by the ITO spheres are weakened. Interestingly, the proposed correction of EMT in Eq. (3) can accurately describe the situation for such 3D models. Based on the field-distribution, we find out the correction factor is 0.66. Then, we calculate the absorptance via the corrected EMT according to Eq. (3), and plot the results as dashed lines in Fig. 5(c), showing a good match with the actual composite simulation results. Moreover, we plot the transmittance on a log scale (i.e. $\log(T)$) through the effective media based on the traditional EMT (black lines) and the corrected EMT (red lines) with $N$ layers of unit cells at the wavelength of 1400nm, as displayed in Fig. 5(d). The interference of multiple reflections is omitted as the total thickness (~300μm) is much larger than the working wavelength. It is seen that the transmittance predicted by the traditional EMT decrease much faster than that predicted by the corrected EMT.

## 5. DISCUSSION AND CONCLUSION

It is noteworthy that the physical mechanism of the breakdown of EMT in 1D dielectric multilayers [16-27] and 2D/3D dielectric composite structures studied here is fundamentally different. In 1D dielectric multilayers, the EMT breaks down close to the total internal reflection angle originates from tunneling effects of evanescent waves. Under the critical incident angle, waves become evanescent in low-$\varepsilon$ layers, but remain propagating in high-$\varepsilon$ layers. Since the layers are deep-subwavelength, the incident waves may still propagate through the multilayer via tunneling, whereas the EMT does not capture this physics, thus leading to the failure of the EMT [16, 17]. Nevertheless, in our proposed 2D/3D structures, the fundamental origin for the breakdown of the EMT is the dramatically varying evanescent fields induced by the field matching condition on the surfaces of the inhomogeneities. When there are tiny absorptive inclusions experiencing such varying local fields, the macroscopic properties (e.g. reflection, transmission and absorption) of the composite structures become sensitive to the positions of those tiny inclusions. Such varying local fields can be comprehended as the excitation of high-order modes [32] instead of dipole-dipole interaction [33]. However, in the traditional EMT description, these details are averaged out, thus leading to the breakdown of the EMT We note that the breakdown in the 2D/3D structures does not rely on the angle of incidence, which can be observed even under normal incidence as shown above.

The continuity of the electric displacement at the inclusion-host interfaces plays an important role in generating the dramatically varying evanescent fields at this deep-subwavelength scale. For instance, in 3D models, the electric fields inside inclusions are nearly uniform. Since the inclusion-host interface is spherical, it is parallel to the fields at some places, but perpendicular to them at other places. This leads to rapidly varying fields in the host nearby the inclusion-host interface. This scenario can also be seen in 2D models for the TM polarization, as we have demonstrated above. However, we also note that for transverse-electric (TE, out-of-plane electric fields) polarization in 2D models, the electric fields nearby the inclusion-host interface are always parallel to the interface. In this case, the traditional Maxwell Garnett EMT is still valid (see details in Supplement 1).

Alternatively, the breakdown of the EMT can also be understood from the mode interactions. In the deep-subwavelength scale, the dipole mode of particles dominates, while high-order modes are generally negligibly small. For subwavelength objects far from each other, the dipole approximation is valid, which is the basis of the traditional EMT [2]. But for objects close to each other, the contribution from high-order modes would be dramatic [32], thus leading to the failure of the traditional EMT.

We note that the discovered position-dependent transmission/absorption characteristics at the deep-subwavelength scale are beyond the extended EMT [34]. The extended EMT are usually used to deal with the composite with components not that subwavelength, where high-order terms are needed to be considered. However, in our deep-subwavelength structures, the high-order terms would be negligibly small. In such deep-subwavelength models, the extended EMT will be consistent with the traditional EMT.

Finally, it is also worth noting that for simple structures like cylinders and spheres, the correction factor $\beta$ can be analytically evaluated based on Eqs. (2) and (4) without using numerical simulations, which lead to $\beta$ in the range of 0~2 (or 0~1.5) for the 2D cylindrical (or 3D spherical) models. Based on these range of values, we can immediately obtain the range of potential deviation for the EMT estimation. This is valuable in many situations. For example, in complicated structures (e.g. Figs. 3 and 4), although numerical analysis is needed to precisely compute $\beta$ as there are no longer simple analytical solutions, our theory can still serve as a guide to design and control absorption without changing the composition of the material. This is unimaginable from the viewpoint of traditional EMT, which is deeply believed in optics. We believe that the numerical simulation won't ruin the value and universality of our proposed corrected EMT.

In summary, we have considered the model of dielectric composites with absorptive constituents. Such a description is generally valid in many circumstances where the absorption is mainly induced by some tiny particles or molecules in the system. Because the region of absorption is determined by the positions of the absorptive constituents, such a case can maximize the difference induced by the evanescent fields at the deep-subwavelength scale. The breakdown of the traditional EMT is inevitable because it simply averages out the evanescent fields and ignores their feature. A correction by taking the distribution of evanescent fields into consideration can significantly increase the accuracy of the EMT prediction. Although traditional wisdom tells us that dielectric structures at the deep-subwavelength scale can be well predicted by the EMT based on homogenized fields, our finding reveals an intriguing exceptional case where the traditional EMT fundamentally breaks down. We have demonstrated that microscopic variation of dielectric structure at the deep-subwavelength scale can also lead to dramatic difference in bulk behaviors, even when the composition of the composite is fixed. Actually, the proposed configuration is common in biomedicine and molecular biology. Nanoparticles are usually used to defect biomolecules and diagnostic assay, forming the configuration consisting of large nanoparticle and nearby absorptive biomolecules [35, 36]. Therefore, we believe that beside the importance in the understanding of the EMT, our work will also be of real interest and impact to advanced photonics, bioscience and practical applications.

**Funding.** National Key R&D Program of China (2017YFA0303702), National Natural Science Foundation of China (11704271, 61671314, 11974176), Natural Science Foundation of Jiangsu Province (BK20170326), the Priority Academic Program Development (PAPD) of Jiangsu Higher Education Institutions.

**Disclosures.** The authors declare no conflicts of interest.

See Supplement 1 for supporting content.

# Supplemental Materials for
## Breakdown of Maxwell Garnett theory due to evanescent fields at deep-subwavelength scale

TING DONG,[1] JIE LUO[2,4], HONGCHEN CHU[1], XIANG XIONG[1], AND YUN LAI[1,3]

[1]*National Laboratory of Solid State Microstructures, School of Physics, and Collaborative Innovation Center of Advanced Microstructures, Nanjing University, Nanjing 210093, China*
[2]*School of Physical Science and Technology, Soochow University, Suzhou 215006, China*
[3]*e-mail: laiyun@nju.edu.cn*
[4]*e-mail: luojie@suda.edu.cn*

1. **Derivation of the correction in EMT**
2. **Calculation of wave transmittance**
3. **Influence of inclusion-size-contrast on the breakdown of Maxwell Garnett theory**
4. **Validity of Maxwell Garnett EMT in 2D models for TE polarization**
5. **Relative error analysis**
6. **Validity check of the simulation results based on COMSOL**
7. **References**

# 1. Derivation of the correction in EMT

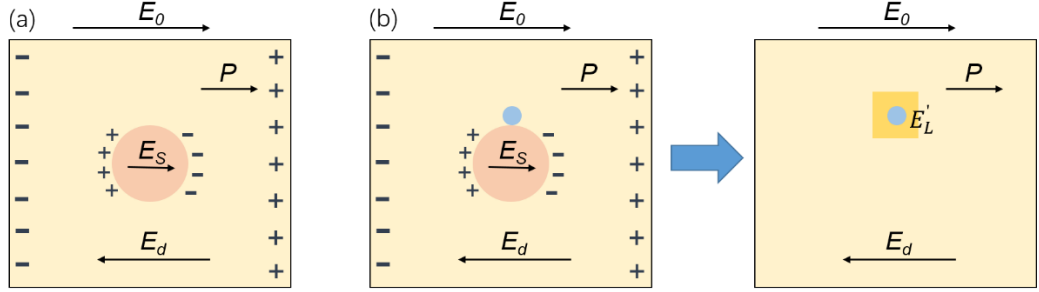

Figure S1. (a) The traditional Lorentz concept for definition of the local fields. (b) The corrected model when a tiny additional inclusion is placed close to the original large inclusion.

Figure S1(a) shows a typical $d$-dimensional ($d=2$ for 2D models, $d=3$ for 3D models) model consisting of a dielectric inclusion in free space in deep subwavelength scale. In this model, the Lorentz local field acting on the inclusion can be decomposed into four components [1, 2]:

$$\mathbf{E}_L = \mathbf{E}_0 + \mathbf{E}_d + \mathbf{E}_s + \mathbf{E}_{near}, \qquad (S1)$$

where $\mathbf{E}_0$ represents the external field, $\mathbf{E}_d$ is the depolarization field due to the bound charges on the outer surface of the host. The sum of $\mathbf{E}_0$ and $\mathbf{E}_d$ is the macroscopic field $\mathbf{E}$ (the homogeneous field averaged over the entire volume of the material), i.e. $\mathbf{E} = \mathbf{E}_0 + \mathbf{E}_d$. $\mathbf{E}_s$ denotes the field due to bound charges on the surface of the sphere, which can be derived as $\mathbf{E}_s = \dfrac{\mathbf{P}}{d\varepsilon_0}$ according to the Coulomb's law. Here, $\mathbf{P}$ is the macroscopic polarization. $\mathbf{E}_{near}$ is the field induced by other inclusions, which vanishes in a cubic/square crystal lattice due to the lattice symmetry. Thus, we can now write the total Lorentz local field as,

$$\mathbf{E}_L = \mathbf{E} + \frac{\mathbf{P}}{d\varepsilon_0}. \qquad (S2)$$

The macroscopic polarization $\mathbf{P}$ can be connected to the local field $\mathbf{E}_L$ via the polarizability $\alpha$ of the inclusion as,

$$\mathbf{P} = N\alpha \mathbf{E}_L = N\alpha\left(\mathbf{E} + \frac{\mathbf{P}}{d\varepsilon_0}\right). \qquad (S3)$$

where $N$ is the volume density of the inclusion.

Combining with the basic constitutive relation $\mathbf{D} = \varepsilon_0\mathbf{E} + \mathbf{P} = \varepsilon_0(1+\chi)\mathbf{E} = \varepsilon_0\varepsilon\mathbf{E}$ ($\chi = \mathbf{P}/\varepsilon_0 \mathbf{E}$ is the dielectric susceptibility), we can build the connection between the polarizability $\alpha$ and the relative permittivity $\varepsilon$ of the medium:

$$\frac{N\alpha}{d\varepsilon_0} = \frac{\varepsilon - 1}{\varepsilon + (d-1)}. \qquad (S4)$$

This is the explicit relation between the microscopic parameter $\alpha$ and the macroscopic observable $\varepsilon$, known as the Clausius-Mossotti relation. Considering the relative permittivity the host $\varepsilon_h$ and the macroscopic observable, i.e. the effective permittivity $\varepsilon_{eff}$, Eq. (S4) turns to be,

$$\frac{N\alpha}{d\varepsilon_0 \varepsilon_h} = \frac{\varepsilon_{eff} - \varepsilon_h}{\varepsilon_{eff} + (d-1)\varepsilon_h}, \qquad (S5)$$

and the polarizability $\alpha$ can be expressed as,

$$\alpha = \frac{d\varepsilon_0 \varepsilon_h f}{N} \frac{\varepsilon_1 - \varepsilon_h}{\varepsilon_1 + (d-1)\varepsilon_h} \qquad (S6)$$

with $f$ being the filling ratio of the inclusion. By substituting Eq. (S6) into Eq. (S5), we obtain,

$$\frac{\varepsilon_{eff} - \varepsilon_h}{\varepsilon_{eff} + (d-1)\varepsilon_h} = f \frac{\varepsilon_1 - \varepsilon_h}{\varepsilon_1 + (d-1)\varepsilon_h}. \qquad (S7)$$

When there exist different inclusions inside the host, Eq. (S7) can be generalized as,

$$\frac{\varepsilon_{eff} - \varepsilon_h}{\varepsilon_{eff} + (d-1)\varepsilon_h} = \sum_i f_i \frac{\varepsilon_i - \varepsilon_h}{\varepsilon_i + (d-1)\varepsilon_h}, \qquad (S8)$$

where $\varepsilon_i$ and $f_i$ are the relative permittivity and filling ratio of the $i$-th inclusion.

Equation (S8) is known as the classical Maxwell Garnett formula, which describes the bulk effective permittivity of the composite in terms of the permittivities and filling fractions of the inclusions of the composite material.

Now, we assume an additional tiny inclusion placed close to the original inclusion, as illustrated in Fig. S1(b). Since the scattering waves by the original inclusion contains a large amount of evanescent waves, dramatically changing electric fields nearby the inclusion. As a result, the additional tiny inclusion would experience varying local fields, which can be much larger or smaller than the average fields in the whole composite. In this case, we can also derive the effective permittivity of this composite following the above methods, but the polarizability $\alpha$ of the tiny inclusion needs correction according to the changed local fields.

We assume that the tiny inclusion is much smaller than the original inclusion, so that the macroscopic field $\mathbf{E}$ and polarization $\mathbf{P}$ of the whole composite are almost unchanged. But locally, the macroscopic field and polarization experienced by the tiny inclusion are changed to $\mathbf{E}' = \beta\mathbf{E}$ and $\mathbf{P}' = \beta\mathbf{P}$, where $\beta$ is the correction factor. Then, the local field experienced by the tiny inclusion turns to be,

$$\mathbf{E}'_L = \mathbf{E}' + \frac{\mathbf{P}'}{d\varepsilon_0} = \beta\mathbf{E}_L. \qquad (S9)$$

Here, the corrected polarization can also be expressed as $\mathbf{P}' = N\alpha'\mathbf{E}_L$, where $\alpha'$ is the corrected polarizability regarding to the original local field $\mathbf{E}_L$. Thus, the dielectric susceptibility of the tiny inclusion regarding to the original polarization $\mathbf{P}$ is,

$$\chi' = \frac{\mathbf{P}}{\varepsilon_0 \mathbf{E}'} = \frac{1}{\varepsilon_0} \frac{\beta^{-1} N\alpha'}{\beta - \frac{N\alpha'}{d\varepsilon_0}}. \qquad (S10)$$

According to the relation $1 + \chi' = \varepsilon_a$ ($\varepsilon_a$ is the relative permittivity of the additional tiny inclusion), we obtain the corrected polarizability as,

$$\alpha' = \frac{d\varepsilon_0}{N} \frac{\beta^2(\varepsilon_a - 1)}{\beta(\varepsilon_a - 1) + d}. \qquad (S11)$$

Considering the relative permittivity the host $\varepsilon_h$ and the filling ratio of the tiny inclusion, the corrected polarizability is rewritten as,

$$\alpha' = \frac{d\varepsilon_0 \varepsilon_h f}{N} \frac{\beta^2(\varepsilon_a - \varepsilon_h)}{\beta(\varepsilon_a - \varepsilon_h) + d\varepsilon_h}. \qquad (S12)$$

In this way, we can correct the effective medium formula in Eq. (S8) as,

$$\frac{\varepsilon_{eff} - \varepsilon_h}{\varepsilon_{eff} + (d-1)\varepsilon_h} = \sum_i^N f_i \frac{\varepsilon_i - \varepsilon_h}{\varepsilon_i + (d-1)\varepsilon_h} + \sum_j^M f_{aj} \frac{\beta_j^2(\varepsilon_{aj} - \varepsilon_h)}{d\varepsilon_h + \beta_j(\varepsilon_{aj} - \varepsilon_h)}, \qquad (S13)$$

where $\varepsilon_i$ and $f_i$ are the relative permittivity and filling ratio of the $i$-th large inclusion. $\beta_j$ denotes the correction factor for the $j$-th additional tiny inclusion (relative permittivity $\varepsilon_{aj}$, filling ratio $f_{aj}$). In numerical calculations, we can evaluate $\beta_j$ by the ratio between the electric field where the $j$-th tiny inclusion lies and the averaged electric field of the whole composite in the absence of all additional inclusions.

## 2. Calculation of wave transmittance

In the analysis of wave transmittance of the effective medium, the interference of multiple reflections is considered when the thickness is comparable with the working wavelength (e.g. Fig. 4), while ignored if the thickness is much larger than the wavelength (e.g. Fig. 3).

The exact transmittance through the effective medium (relative permittivity $\varepsilon_{eff}$, thickness $d$) in a background medium with relative permittivity of $\varepsilon_b$ can be calculated as,

$$T = \left| \frac{tt'e^{i\delta}}{1 - r^2 e^{i2\delta}} \right| \quad \text{(S14)}$$

where $t = \frac{2\sqrt{\varepsilon_b}}{\sqrt{\varepsilon_b} + \sqrt{\varepsilon_{eff}}}$, $t' = \frac{2\sqrt{\varepsilon_{eff}}}{\sqrt{\varepsilon_b} + \sqrt{\varepsilon_{eff}}}$, $r = \frac{\sqrt{\varepsilon_b} - \sqrt{\varepsilon_{eff}}}{\sqrt{\varepsilon_b} + \sqrt{\varepsilon_{eff}}}$ and $\delta = \sqrt{\varepsilon_{eff}} k_0 d$ for normal incidence. $k_0$ is the wave number in free space. In Eq. (S14), the interference of multiple reflections is considered. Usually, when the thickness of the effective medium is much larger than the working wavelength, we can ignore the interference, thus the transmittance reduces to,

$$T = \left| tt'e^{i\delta} \right|. \quad \text{(S15)}$$

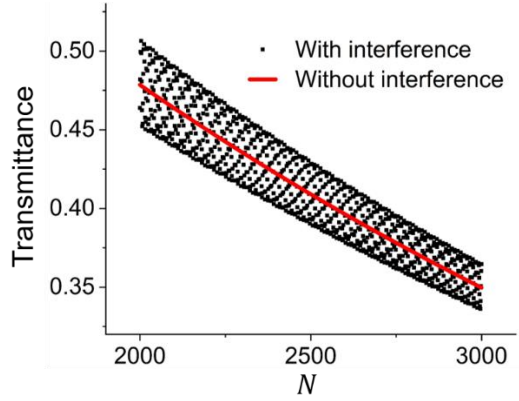

Figure S2. Transmittance through the effective medium slab in Fig. 4 based on Eq. (S14) (black dots) and Eq. (S15) (red lines) as a function of the number of unit cells.

We take the 3D sample in Fig. 4 to compare the transmittance of the two cases. Figure S2 plots the transmittance based on Eq. (S14) (black dots) and Eq. (S15) (red lines) as a function of the number of unit cells. We see that the transmittance based on Eq. (S15) always lies between the maximal and minimal transmittance based on Eq. (S14). This indicates that the transmittance without considering the interference of multiple reflections can still well describe the transmission characteristics of the effective medium.

## 3. Influence of inclusion-size-contrast on the breakdown of Maxwell Garnett theory

The evanescent-wave-induced varying fields are localized nearby the inclusions, and smoothed at a distance comparable to the size of the inclusions. Therefore, the size-contrast of inclusions is crucial in the observation of EMT breakdown. Only when additional inclusions are very small compared with the large inclusions, they can "see" the varying local fields instead of averaged fields. If we gradually increase the size of the additional inclusions, the correction factor will tend to be unity, and the corrected EMT finally becomes the traditional EMT.

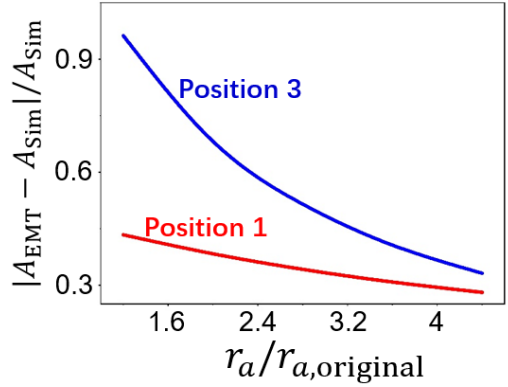

Figure S3. The $|A_{\mathrm{EMT}} - A_{\mathrm{Sim}}|/A_{\mathrm{Sim}}$ as a function of the radius of the additional tiny inclusion at positions 1 (red lines) and 3 (blue lines). The model is the same as that in Fig. 2(a) in the Main Text, except that the radius of the additional tiny inclusion (original radius $r_{a,\mathrm{original}}$) is changed.

For demonstration, we take the model in Fig. 2(a) as an example. We gradually increase the radius of the additional tiny inclusion to 4.4 times of the original one (other parameters are unchanged). We find that the correction factor $\beta$ at position 1 (or 3) is gradually decreased (or increased) from is 1.386 (or 0.7165) to 1.214 (or 0.8929). Moreover, we calculate the $|A_{\mathrm{EMT}} - A_{\mathrm{Sim}}|/A_{\mathrm{Sim}}$ as the function of the radius of the additional tiny inclusion at position 1 (red lines) and position 3 (blue lines), as plotted in Fig. S3. $A_{\mathrm{Sim}}$ and $A_{\mathrm{EMT}}$ denote the absorptance of the composite based on simulations of the actual composite and the traditional EMT (i.e. Eq. (1) in the Main Text), respectively. It is clearly seen that as the increase of the size of the additional inclusion, the deviation of the traditional EMT becomes smaller. This indicates that the breakdown of the Maxwell Garnett EMT will become unobvious as the decrease of the size-contrast between inclusions.

## 4. Validity of Maxwell Garnett EMT in 2D models for TE polarization

In the Main Text, we observe the breakdown of Maxwell Garnett EMT in 2D models for transverse-magnetic (TM, out-of-plane magnetic fields) polarization. This is because the continuity conditions of electric field and electric displacement at the inclusion-host interfaces lead to the locally varying fields. However, for transverse-electric (TE, out-of-plane electric fields) polarization, the electric fields nearby the inclusion-host interface are nearly uniform as there are only components parallel to the interface. In this case, the traditional Maxwell Garnett EMT is still valid, and the effective permittivity of a deeply subwavelength 2D composite can be calculated as [1, 2],

$$\varepsilon_{eff} = \left(1 - \sum_i f_i\right)\varepsilon_h + \sum_i f_i \varepsilon_i, \qquad (S16)$$

where $\varepsilon_i$ and $f_i$ are the relative permittivity and filling ratio of the $i$-th inclusion.

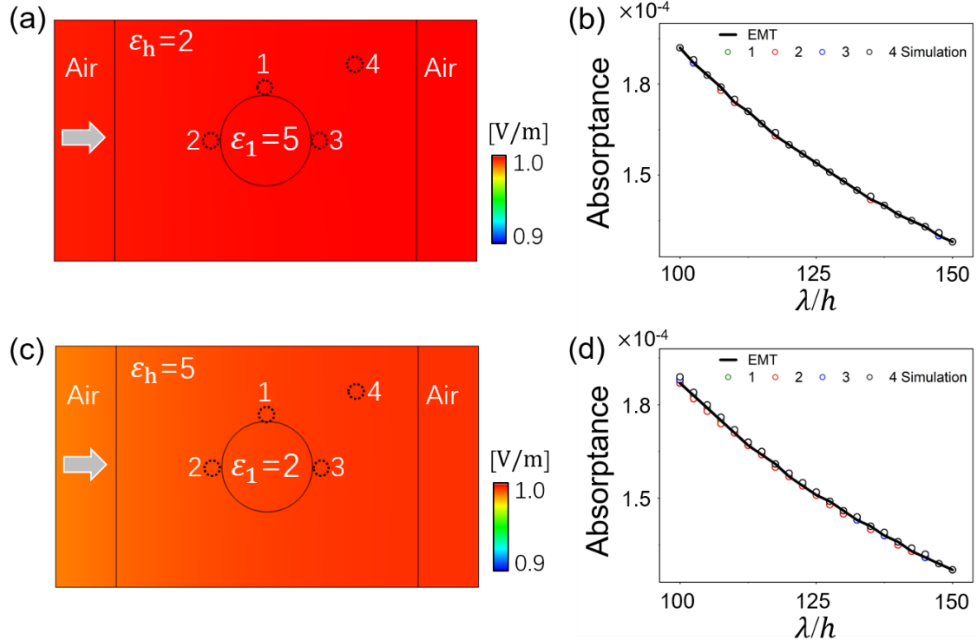

Figure S4. [(a) and (c)] The $|\mathbf{E}|$-distributions for the composite structure with (a) $\varepsilon_h = 2$ and $\varepsilon_1 = 5$, (c) $\varepsilon_h = 5$ and $\varepsilon_1 = 2$ illuminated by a TE-polarized wave under normal incidence. The working wavelength is $\lambda = 125h$. The dashed circles denote the positions of additional tiny inclusions. [(b) and (d)] Absorptance by the composite with (b) $\varepsilon_h = 2$ and $\varepsilon_1 = 5$, (d) $\varepsilon_h = 5$ and $\varepsilon_1 = 2$ as the function of working wavelength based on simulations of the actual composite (dots) and traditional EMT (solid lines) when an additional tiny inclusion successively moves from position 1 to position 4. The model is the same as that in Fig. 2 in the Main Text.

For verification, we take the 2D model in Fig. 2 as an example. We change the polarization of incident to TE polarization, and keep other parameters unchanged. Figure S4(a) and S4(c) present the $|\mathbf{E}|$-distributions for the composite structure with $\varepsilon_h = 2$ and $\varepsilon_h = 5$, respectively. It is clearly seen that the electric fields are uniform around the large inclusion. As a result, the traditional Maxwell Garnett EMT is still valid in describing such a composite structure, as confirmed by the absorptance based on simulations (dots) and EMT (solid lines) in Figure S4(b) and S4(d).

## 5. Relative error analysis

Here, we first calculate the relative error $|A_{EMT} - A_{Sim}|/A_{Sim}$ as the function of $\lambda/h$ for the 2D model studied in Fig. 2(a) in the Main Text, as shown in Fig. S5(a). $A_{Sim}$ and $A_{EMT}$ denote the absorptance of the composite based on simulations of the actual composite and the traditional Maxwell-Garnett EMT (i.e. Eq. (1) in the Main Text), respectively. The blue and red lines denote the cases with the additional tiny inclusion placed at position 1 and position 3, respectively. We see that the relative error remains almost unchanged when the $\lambda$ increases as the whole system is in the deep-subwavelength scale ($\lambda/h \approx 125$).

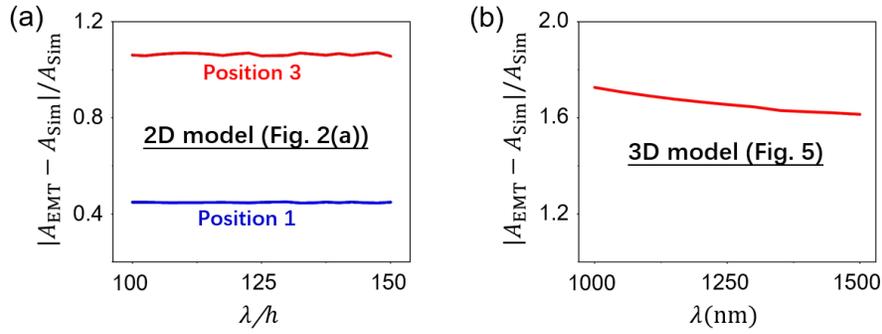

Figure S5. (a) The relative error $|A_{EMT} - A_{Sim}|/A_{Sim}$ as a function of $\lambda/h$ for the 2D model studied in Fig. 2(a) in the Main Text. The blue and red lines denote the cases with the additional tiny inclusion placed at positions 1 and position 3, respectively. (b) The relative error $|A_{EMT} - A_{Sim}|/A_{Sim}$ as a function of $\lambda$ for the 3D model studied in Fig. 5 in the Main Text.

We also calculate the relative error for the 3D model studied in Fig. 5 in the Main Text, as plotted in Fig. S5(b). Since this model is not that subwavelength ($\lambda/h \approx 25$), we find that the relative error decreases as the wavelength increases. This indicates that the Maxwell-Garnett EMT would usually fail less as the wavelength increases (like the 3D model) until the relative error reaches a stable value (like the 2D model). This is because the evanescent fields won't change with $\lambda$ when the $\lambda$ is increased to a certain extent. Under this circumstance, the evanescent fields are determined by the field continuity condition under the quasi-static limit (Eqs. (2) and (4) in the Main Text).

## 6. Validity check of the simulation results based on COMSOL

In order to ensure more rigorous results, here we recalculate the 3D model in Fig. 5 in the Main Text by using FDTD software CST, as plotted in Fig. S6. We see that the absorptance based on COMSOL (red dots) perfectly matches that from CST (black solid lines), confirming the validity of the simulated results.

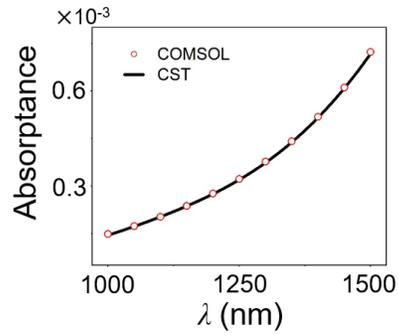

Figure S6. Absorptance of the 3D model in Fig. 5 in the Main Text calculated based on finite-element software COMSOL (red dots) and FDTD software CST (black solid lines).